\documentclass[a4paper,10pt]{article}
\usepackage{amsmath}

\begin{document}

\title {On the spectrum and eigenfunctions of the equivariant general
boundary value problem outside the ball for the
Schr\"{o}dinger operator
with Coulomb potential}

\author{V.P. Burskii, A. A. Zaretskaya}

\date{\today}

\begin{abstract}

We consider the Schr\"{o}dinger equation for hydrogen-like atom with Coulomb potential and non-point ball nucleus.
The eigenvalues and eigenfunctions of the operator given by an arbitrary rotation-invariant boundary value
problem on the spherical boundary of the nucleus are found and as it is proved to be the eigenvalues
are independent on selection of any such boundary value problem and they are the same as for point nucleus.

\end{abstract}

\maketitle

We consider the Schr\"{o}dinger equation for hydrogen-like atom with Coulomb potential and non-point ball nucleus.
The eigenvalues and eigenfunctions of the operator given by an arbitrary rotation-invariant boundary value
problem on the spherical boundary of the nucleus are found and as it is proved to be the eigenvalues
are independent on selection of any such boundary value problem and they are the same as for point nucleus, i.e. the spectrum of radiation and absorption of the hydrogen atom does not change when the size of the nucleus changes.
Interest in the study of the influence of the size of the nucleus on quantum-mechanical parameters is associated, in particular, with the appearance of an article on experiments with muonic hydrogen. In 2010, article 32 of the authors published results on data that indicate that the size of the proton is 4 percent smaller than it follows from calculations of quantum electrodynamics
("The size of the proton" http://www.nature.com/nature/journal/v466/n7303/full/nature09250. html).


In this paper the emission (absorption) spectrum of a hydrogen-like atom
with nontrivial nucleus radius was found, it is understood as the discrete
spectrum of the operator in the title. Assume that the nuclear charge is distributed spherically symmetric.
As is well known, spherically symmetric body beyond its limits creates
the same gravitational field, as a material point of the same mass,
which is located in the center of the body. Therefore we use the Coulomb potential,
using the analogy with gravity. Usually two restrictions are imposed
on the wave function, that is a solution of the Schr\"odinger equation with
the Coulomb potential, they are a limitation at zero and the decrease
at infinity \cite{7}. In present paper, the wave function is not defined
in a neighborhood of zero, instead it we consider a boundary value problem
for solution in the exterior of a sphere of radius $ \rho_0 $.
We do not know what the boundary conditions should be placed on the surface
of the nucleus, but we assume that they must be spherically symmetric.
This leads to the formulation of the general equivariant boundary value problem.
In this paper we consider the general external rotation-invariant boundary value
problem for the Schr\"odinger equation with the Coulomb potential.
The eigenvalues and the corresponding eigenfunctions of the problem
were obtained. It is proved to be that obtained energy values are the same as radiation
energy of the point size atom, that sounds awesome, although, of course, the eigenfunctions are other.
The Schr\"odinger equation is usually studied in the whole space, boundary value problems for the Schr\"odinger type equation have been studied in some papers, for instant in the works \cite{1,2,9}, but the setting as above has not been considered. Remark that arbitrary rotation-invariant boundary value problems for the PDEs have been
considered in the book \cite{4}.



Let us consider the stationary Schr\"{o}dinger equation for the wave
function of an electron of mass $ M $ and the Coulomb attractive potential
in the exterior of the ball $K=\{x\in \textbf{R}^3, |x|<\rho_0\}$ with
a general boundary value problem :
\begin{equation}
\label{1}
\left(\triangle_{x,y,z}+\frac{2M}{\hbar^2}\left(\frac{Ze^2}{r}+E\right)\right)\psi(r,\varphi,\theta)=0,
\end{equation}
\begin{equation}
\label{2}
A\psi|_{\partial K}+B\psi'_\nu|_{\partial K}=0.
\end{equation}
Here $-\frac{Ze^2}{r}$ -- potential,
$e$ -- electron charge, $Ze$ -- nucleus charge,
$E$ --    eigenvalue,
$\hbar$ --  Dirac constant,
$\psi(r,\varphi,\theta)$ -- unknown wave function.
We assume that the boundary value problem (\ref {2}) with normal $\nu$ is
invariant with respect to ball rotations that is the operators
$A$ and $B$ are invariant.

Let's consider the quasi-regular unitary representation
\newline $T~:~G~\rightarrow~U(L_2(S^{2})),$
$\left[T(g)f\right](\xi)=f(g^{-1}\xi),\,$
$f(\xi)\in L_2(S^{2}),\, g\in G
$  of the Lie group $G=SO(3)$.
It is well-known \cite{6} that every linear operator in
$ L_2 (S ^ {2}) $ which commutes with all operators $ T (g) $ of
quasi-regular representation is convolutional.
Therefore we will  consider boundary problems of the form
\begin{equation}
\label{3}
\psi|_{\partial K}*\alpha+\psi'_\nu|_{\partial K}*\beta=0,\quad \alpha^2+\beta^2\neq 0.
\end{equation}
Here $\alpha$ and $\beta$ are arbitrary given functions on the sphere $\partial K$. At infinity we have set the condition of disappearance.
We want to find the eigenvalues of operator from (\ref{1}) with condition
(\ref{2}) and show that these eigenvalues don't depend on functions $\alpha$ and $\beta$.

For investigation of this problem we will use the well-known way specified in the standard books \cite{8,11}. It appears that the method of separation of variables is also suitable in this case of Sch\"odinger equation with the general boundary value problem. First, let's write the general solution of equation (1).
Suppose that the solution in polar coordinates is represented
in the form
\begin{equation}
\label{4}
\psi(r,\varphi,\theta)=\hat{C}\sum_{l=0}^\infty\sum_{m=-l}^l R_{\,l,\,m}(r)Y_{l,m}(\varphi,\theta),
\end{equation}
where 
$Y_{l,m}=\frac{1}{\sqrt{2 \pi}} e^{i m \varphi}(-1)^m \sqrt{\frac{2l+1}{2} \frac{(l-m)!}{(l+m)!}}P^m_l (\cos\theta)$ are spherical functions that are eigenfunctions of the square
of the angular momentum with eigenvalues $l(l+1), l=0,1,2..\infty$, $P^m_l (\cos\theta)$
are associated Legendre functions, $\hat{C}$ is a constant, which is convenient
for us to enter at once but choose it later and $R_{\,l,\,m}(r)$ are unknown radial functions.

Substituting (\ref {4}) into (\ref {1}) we
obtain the following equation for the radial parts of the wave function
\begin{eqnarray}
&&R_{l,m}''+\frac{2}{r}R_{l,m}'+\nonumber\\
\label{5}&&+R_{l,m}\left(-l(l+1)\frac{1}{r^2}+\frac{2MZe^2}{\hbar^2}\frac{1}{r}+\frac{2ME}{\hbar^2}\right)=0.
\end{eqnarray}

Let's find  a solution of equation (5) explicitly by
ma\-king the change
$ R_{\,l,\,m}(r)=\widehat{R}_{\,l,\,m}(\rho)\rho\,^l\,e^{-\rho/2},\ \ \rho=2nr $.

For convenience we introduce the notation
\begin{equation*}
\sqrt{-2ME}\,/\,\hbar=n
\end{equation*}
considering the case $ E<0 $.

Then we obtain the following equation
\begin{equation}
\label{6}
\rho\widehat{R}_{\,l,\,m}''+(2l+2-\rho)\widehat{R}_{\,l,\,m}'+
\widehat{R}_{\,l,\,m}(\frac{Me^2}{n\hbar^2}-l+1)=0.
\end{equation}
The last equation is degenerate hypergeometric equation and solutions are the Kummer functions
with the first parameter $-\frac{Me^2}{n\hbar^2}+l-1$ and with the second parameter $2l+2$.
Therefore, in terms of the degenerate hypergeometric functions
of the first and second kinds, we get \cite{3}
\begin{eqnarray}\label{7}
&&\widehat{R}_{\,l,\,m}(\rho)=C_1(l,m)\rho^{l}e^{-\rho/2}\Phi(l-\frac{Me^2}{n\hbar^2}-1,2l+2,\rho)+\nonumber\\
&&\label{10}+C_2(l,m)\rho^{l}e^{-\rho/2}\rho^{-2l-1}\Psi(-l-\frac{Me^2}{n\hbar^2}-2,-2l,\rho).
\end{eqnarray}

Note that the function $\widehat{R}_{\,l,m}$ also depend on $n$ (further on $k$).
Let's investigate the behavior of the radial part of the wave function
at infinity using equation \eqref{5}. Let $r$ take large values,
then some terms can be neglected in equation \eqref{5}, namely
those which are multiplied by $\frac{1}{r}$ or
$\frac{1}{r^2}$.
We obtain the equation
$R^{\prime\prime}+\frac{2ME}{\hbar^2}R=0 $.
It has a finite solution at infinity $R=e^{-nr} $.
Hence, the solution of equation \eqref{5} at infinity should decrease
as $e^{-nr} $. It means that the function
$ \Phi(-\frac{Me^2}{n\hbar^2}+l-1, 2l+2,\rho)$
should not grow  at infinity too fast.
However, it is well-known\cite{3} that generic
degenerate hypergeometric functions increases as the exponent of its argument.

In order to the degenerate hypergeometric function of the first kind
in \eqref{7} does not spoil the behavior of the radial function at infinity, it is necessary that the first parameter would be a negative integer.
\begin{eqnarray}\label{8}
\Phi(\alpha,\beta,z)=1+\frac{\alpha}{\beta}\frac{z}{1!}&+&
\frac{\alpha(\alpha+1)}{\beta(\beta+1)}\frac{z^2}{2!}+\nonumber\\
\label{12}&+&\frac{\alpha(\alpha+1)(\alpha+2)}{\beta(\beta+1)(\beta+2)}\frac{z^3}{3!}+...
\end{eqnarray}
As it can be seen from the definition of the
degenerate hypergeometric function \eqref{8}, if the first parameter
$ \alpha $ is an integer negative then all
the terms in the series will be nulled, except for the first some terms.
Thus, the function $\Phi\ $ from \eqref{8} becomes a polynomial
function and hence the corresponding term in \eqref{8} disappearances at infinity. Denote a negative integer value of the parameter $\alpha$
in \eqref{8} by $-\frac{Me^2}{n\hbar^2}+l-1=-k+l-1$.
It is clearly that the parameter $ k $ (known as the principal quantum
number) can be any positive integer, $k\geq l-1, l\geq0$.
So, $n=\frac{Me^2}{\hbar^2k}$.

Let's see how the function $\Psi(-k-l-2,-2l,x) $ behaves at infinity\cite{3}.
\begin{eqnarray}
\label{13}
\Psi(-k-l-2,-2l,\rho)\approx\qquad \nonumber\\
\approx\sum_{p=0}^N(-1)^p \frac{(-l-k-2)_p(l-k-1)_p}{p\,!}
\rho^{l+k+2-p}+
\\
\nonumber
+O(|x|^{l+k+2-N-1}).
\end{eqnarray}
From the last equation it is clear that the function at infinity behaves
as a polynomial.

Returning to \eqref{6} we can see that the energy values are
\begin{equation}
\label{15}
E_{k}=\frac{-Me^4}{2\hbar^2k^2}, \,\,k=\overline{1..\infty}.
\end{equation}
\emph{
Note that in this case the energy levels is the same as in the classical case of a point nucleus.
}

Return to the general boundary value problem \eqref{3}:
$\psi|_{\partial K}*\alpha+\psi'_\nu|_{\partial K}*\beta=0.$
Let $\alpha=\sum_{l=0}^\infty \sum_{m=-l}^l \alpha_l^mY_{l,m}$,
$\beta=\sum_{l=0}^\infty \sum_{m=-l}^l   \beta_l^mY_{l,m}$ --
decompositions of functions in Fourier series on sphere $S^2$.
The function $\psi$ from \eqref{4} depend on $E$ or on $k$ by virtue of \eqref{15}. For different $k$ we have different eigenfunctions $\psi_k$.
For the eigenfunction $\psi_k$ also
$\psi_k|_{\partial K}= \sum_{l=0}^\infty \sum_{m=-l}^l a_{k,\,l}^mY_{l,m}$,
${\psi_k'}_\nu|_{\partial K}= \sum_{l=0}^\infty \sum_{m=-l}^l b_{k,\,l}^mY_{l,m}$.
$*$ --  convolution on $\partial K$,
 that is $$\psi_k|_{\partial K}*\alpha=\sum_{l=0}^\infty \sum_{m=-l}^l
a_{k,\,l}^m\alpha_l^0 Y_{l,\,m}.$$

Here we are in the following situation and we will use the following results. As usual, an operator $A$ is called invariant with respect to a group $G$ of transformations if the operator $A$ commutes with transformations of $G$, more accurate, if the operator $A$ commutes with
each operator of quasiregular representation of the group $G$ in the action space of $A$. For the space $E=L_2(S^2)$ and the group $G=SO(3,\bf R)$ acting on $S^2$ by rotations a quasiregular representation $T:G\to GL(E)$ is given by definition by the formula\cite{6} $T(g)f(x)=f(g^{-1}x,$ $f(x)\in E,\ g\in G$. This representation is unitary and has the decomposition in the direct sum of irreducible representations $T=\sum_{l=0}^{\infty } T^l,$ where irreducible component $T^l$ acts on the space $\gamma (H^l)$, $\gamma$ is the operator of contraction of functions from $\bf R^3$ onto sphere $S^2$, $H^l$ is the space of homogeneous harmonic polynomials of degree $l$ on $\bf R^3$ and the spherical functions $\{Y_{\,m,\,l}\}_{m=-l}^l $ constitute a basis in $H^l$. We will consider any function on sphere $S^2$ as a function on the group $G$ that is a constant on each left coset in $G/SO(2,\bf R)$. The following statements\cite{12} hold:

\vskip10pt
1). For any linear operator $\mathcal A$ in the space $E$ which commutes with each operator $T(g)$ of quasiregular representation there exists a function $\psi_{\mathcal A} \in L_2(S^{n-1})$ such that
$$
\left[ {\mathcal A} \varphi \right] (g) = \int\limits_{g_1 \in G} \varphi (g_1)
\psi_{\mathcal A}
(g_1^{-1}g)dg_1 = \left[ \varphi * \psi_{\mathcal A} \right] (g)
$$
for each $\varphi \in L_2(S^{n-1})$. Vise verse any convolution operator commutes with each $T(g)$.

\vskip10pt
2). Let us have
$$
f_1 * f_2 (g) = \sum_{l=0}^{\infty} \sum_{m=1}^{h(l)} \gamma_l^m t_{m1}^l (g)
$$
is decomposition of convolution of functions $f_1 (g)$ и $f_2 (g)$
that are constants on left cosets in Fourier expansion.
Here $t_{m1}^l (g) = \left( T^l(g)e_1,e_m \right)$ are matrix entries of
irreducible representation  $T^l$, $e_1$ is an invariant with respect to
$H$ vector in the space $T^l$, $h(l) = \dim T^l$.
Then $\gamma_l^m = \lambda_l^m \cdot \mu_l^1$, where
$\lambda_l^m$ and $\mu_l^m$ are Fourier coefficients of functions
$f_1 (g)$ and $f_2 (g)$ respectively. For our case we have the basis $t^l_{m1}$ corresponds to $Y_{\,l,\,m}$, the index 1 in $\mu_l^1$ means
zonal harmonic $Y_{\,l,\,0}$.

Boundary value problem (\ref {3}) in terms of the Fourier coefficients
for each $ k $ can be written as
\begin{eqnarray}
\label{18}
a_{k,\,l}^m\alpha_l^0+b_{k,\,l}^m\beta_l^0=0.
\end{eqnarray}
$a_{\,k,\,l}^m=\widehat{R}_{k,\,l,\,m}|_{\rho=\rho_0}$ is given by
\eqref{4},\eqref{7} and
$b_{k,\,l}^m=\left.\frac{1}{\rho_0}\frac{\partial \widehat{R}_{k,l,m}(\rho)}{\partial \rho}
\right|_{\rho=\rho_0}.$ Note that tesseral harmonics in $\alpha$ and $\beta$ can be omit.

Thus, the boundary value problem can be written as
\begin{eqnarray}\label{20}
\{C_1(k,l,m)\rho_0^{2l+3}\Phi(l-k-1,2l+2,\rho_0)+\nonumber\\
+C_2(k,l,m)\rho_0^{2}\Psi(-k-l-2,-2l,\rho_0)\}\alpha_l^0+\nonumber\\
+\{C_1(k,l,m)\rho_0^{2l+1}(l\Phi(l-k-1,2l+2,\rho_0)-\nonumber\\
-1/2\rho_0\Phi(l-k-1,2l+2,\rho_0)+\qquad\\
+\rho_0\frac{l-k-1}{2l+2}\Phi(l-k,2l+3,\rho_0))+\quad\nonumber\\
+C_2(k,l,m)((-l-1)\Psi(-k-l-2,-2l,\rho_0)-\nonumber\\
-1/2\rho_0\Psi(-k-l-2,-2l,\rho_0)+\nonumber\\
+\rho_0(l+k+2)\Psi(-k-l-1,-2l+1,\rho_0))\}\beta_l^0=0.\nonumber
\end{eqnarray}

We can assume that $C_2(k,l,m)\equiv 1$, since both sides
of \eqref{20} can be divided into an arbitrary constant.
The equality \eqref{20} allows us to find unknown constant
$C_1(k,l,m)$.

Normalization condition allows us to find the last unknown constant
$\hat{C}$ from \eqref{4}.
\begin{eqnarray}
\int_{r_0}^\infty |\psi_k(r,\varphi,\theta)|^2dr=1.\nonumber
\end{eqnarray}

\newpage
{\bf \large Conclusion.}
\emph{
Eigenvalues and the corresponding eigenfunctions of the problem (1), (2),
in the above notation, are}

\[
E_{k}=\frac{-Me^4}{2\hbar^2k^2}, \,\,k=\overline{1..\infty},
\]
\begin{eqnarray*}
\psi_k(r,\varphi,\theta)=\hat{C}\sum_{l=0}^k\sum_{m=-l}^l
(C_1(k,l,m)(2nr)^{l}e^{-nr}   \nonumber\\\times\Phi(l-k-1,2l+2,2nr)+\nonumber\\
+(2nr)^{-l-1}e^{-nr}\Psi(-l-k-2,-2l,2nr))
Y_{l,m}(\varphi,\theta).
\end{eqnarray*}



\end{document}